\documentclass[11pt,fleqn,twoside]{article}
\usepackage{amsfonts,amssymb,latexsym}
\makeatletter
\newcommand{\prava}{\footnotesize\it
\begin{flushright}
\begin{minipage}{18cm}%{6cm}%9.6
Copyright \copyright 1998 by F. Calogero
\end{minipage}
\end{flushright}}

\newcommand{\name}[1]{\begin{flushleft}
                       \LARGE \bf #1
                       \end{flushleft}\vspace{-3mm}}

\newcommand{\Author}[1]{\begin{flushleft}
                       \it #1 \end{flushleft}}

\newcommand{\Adress}[1]{\begin{flushleft}
                       \it #1 \end{flushleft}}

\newcommand{\Date}[1]{\begin{flushleft}
                      \small  \it #1 \end{flushleft}}

\newcommand{\ehkol}{Author \ name}
\newcommand{\ohkol}{Article \ name}
\renewcommand{\@evenhead}{
\hspace*{-3pt}\raisebox{-15pt}[\headheight][0pt]{\vbox{\hbox to \textwidth
{\thepage \hfil \ehkol}\vskip4pt \hrule}}}
\renewcommand{\@oddhead}{
\hspace*{-3pt}\raisebox{-15pt}[\headheight][0pt]{\vbox{\hbox to \textwidth
{\ohkol \hfil \thepage}\vskip4pt\hrule}}}
\renewcommand{\@evenfoot}{}
\renewcommand{\@oddfoot}{}

\newcommand{\be}{\begin{equation}}
\newcommand{\ee}{\end{equation}}
\newcommand{\ba}{\hspace*{-5pt}\begin{array}}
\newcommand{\ea}{\end{array}}

\newcommand{\ds}{\displaystyle}
\makeatother

\begin{document}
\setcounter{page}{289}

\thispagestyle{empty}

\renewcommand{\ehkol}{F. Calogero}
\renewcommand{\ohkol}{A Solvable Many-Body Problem in the Plane}

\begin{flushleft}
\footnotesize \sf
Journal of Nonlinear Mathematical Physics \qquad 1998, V.5, N~3,
\pageref{calogero-fp}--\pageref{calogero-lp}.
\hfill {\sc Article}
\end{flushleft}

\vspace{-5mm}

\renewcommand{\footnoterule}{}
{\renewcommand{\thefootnote}{}
 \footnote{\prava}}

\name{A Solvable Many-Body Problem in the Plane}\label{calogero-fp}

\Author{F. CALOGERO}

\Adress{$^*$~Dipartimento di Fisica, Universit\`a di Roma ``La Sapienza'',
00153 Roma, Italy\renewcommand{\thefootnote}{}\footnote{$^*$On leave
while serving as Chairman of the Council, Pugwash
Conferences on Science and World Affairs}\\[1mm]
~~Istituto Nazionale di Fisica Nucleare, Sezione di Roma, Italy}

\Date{Received May 1, 1998}

\begin{abstract}
\noindent
A solvable many-body problem in the plane is exhibited. It is
characterized by rotation-invariant Newtonian (``acceleration equal
force'') equations of motion, featuring one-body (``external'')
and pair (``interparticle'') forces. The former depend quadratically
on the velocity, and nonlinearly on the coordinate, of the moving
particle. The latter depend linearly on the coordinate of the moving
particle, and linearly respectively nonlinearly on the velocity
respectively the coordinate of the other particle. The model contains
$2n^2$ arbitrary coupling constants, $n$ being the number of
particles. The behaviour of the solutions is outlined; special cases
in which the motion is conf\/ined (multiply periodic), or even
completely periodic, are identif\/ied.
\end{abstract}

\setcounter{equation}{0}
\renewcommand{\theequation}{\arabic{section}.\arabic{equation}}

\section{Introduction}

Recently several solvable and/or integrable many-body problems in the
plane have been introduced [1]--[4]. (We use here the heuristic --
imprecise but useful -- distinction among {\it solvable} models, for
which the solutions can be obtained in relatively explicit form, and
{\it integrable} models, which possess a suf\/f\/iciently large number of
independent globally def\/ined constants of the motion). In this paper
we report one more such model, which seems worthy of special notice
because of its simplicity and generality ($2n^2$ arbitrary coupling
constants, see below). Indeed its equations of motion read as
follows:
\be
\ba{l}
\ds \ddot{\vec r_j}= \frac{ 2\dot{\vec r_j}(\dot{\vec r_j}\cdot {\vec
r_j}) -{\vec r_j}(\dot{\vec r_j}\cdot \dot {\vec r_j})}{r_j^2}\\[5mm]
\ds \qquad \qquad \qquad +\sum_{k=1}^n (\beta_{jk}+\gamma_{jk}
\hat{z} \wedge) \frac{\dot{\vec r_k}({\vec r_k}\cdot {\vec r_j}) +
{\vec r_j}(\dot{\vec r_k}\cdot {\vec r_k})-
{\vec r_k}(\dot{\vec r_k}\cdot {\vec r_j})}{r_k^2}.
\ea
\ee

\noindent
{\bf Notation.} Superimposed arrows identify 2-vectors, for which we
use the notation
\be
\vec r \equiv (x,y), \qquad \hat{z} \wedge \vec r\equiv (-y,x), \qquad
r^2=x^2+y^2.
\ee
(The second formula corresponds to the 3-dimensional notation $\vec
r\equiv (x,y,0)$, $\hat z\equiv (1,0,0)$, with the symbol $\wedge$
denoting the standard 3-dimensional vector product). The index $j$
always takes integer values from 1 to $n$, the total number of
interacting particles. The $2n^2$ ``coupling constants'' $\beta_{jk}$
and $\gamma_{jk}$ are arbitrary. Dots denote of course
dif\/ferentiation with respect to the time $t$.

\medskip

The equations of motion (1.1) are of Newtonian type (``acceleration
equal force''), with one- and two-body forces. The one-body
(``external'') force is quadratic in the particle velocity $\dot{\vec
r_j}$, and depends nonlinearly on the particle coordinate $\vec r_j$.
The two-body (``interparticle'') force depends linearly (only) on the
coordinate of the moving particle, and linearly respectively
nonlinearly on the velocity respectively the coordinate of the other
particle. The model is {\it rotation-invariant}, but not {\it
translation-invariant} (see however below for a {\it
translation-invariant} version). It is invariant under the {\it scale
transformation} $\vec r_j \to c \vec r_j$, with $c$ an arbitrary
constant.

This model is {\it solvable}, as explained in the following Section.
In the subsequent Section~3 a brief discussion is given of the
particle motions it entails, and the following generalized version of
(1.1) is also introduced:
\setcounter{equation}{2}
\renewcommand{\theequation}{\arabic{section}.\arabic{equation}{\rm a}}
\be
\ba{l}
\ds \ddot{\vec r_j}= (\lambda +\omega \hat z \wedge)\dot{\vec r_j}+
\frac{ 2\dot{\vec r_j}(\dot{\vec r_j}\cdot {\vec
r_j}) -{\vec r_j}(\dot{\vec r_j}\cdot \dot {\vec r_j})}{r_j^2}\\[5mm]
\ds \qquad \qquad \qquad +\sum_{k=1}^n (\widetilde{\beta}_{jk}(t)+
\widetilde{\gamma}_{jk}(t)
\hat{z} \wedge) \frac{\dot{\vec r_k}({\vec r_k}\cdot {\vec r_j}) +
{\vec r_j}(\dot{\vec r_k}\cdot {\vec r_k})-
{\vec r_k}(\dot{\vec r_k}\cdot {\vec r_j})}{r_k^2},
\ea
\ee
\setcounter{equation}{2}
\renewcommand{\theequation}{\arabic{section}.\arabic{equation}{\rm b}}
\be
\ba{l}
\ds \widetilde{\beta}_{jk}(t) = \left[ \beta_{jk} \cos(\omega t)-
\gamma_{jk}\sin(\omega t)\right] \exp(\lambda t),\\[2mm]
\ds \widetilde{\gamma}_{jk}(t) = \left[ \gamma_{jk} \cos(\omega t)+
\beta_{jk}\sin(\omega t)\right] \exp(\lambda t).
\ea
\ee
For $\lambda=\omega=0$, this generalized model reduces to the
previous one; for $\lambda=0$, $\omega\not= 0$ (and of course {\it
real}), {\it all} its solutions are completely periodic, with period
$T=2\pi/\omega$ (see Section~3). In Section~3 we also indicate how to
manufacture analogous solvable models, which are also {\it
translation-invariant}.

Let us f\/inally emphasize that many-body models {\it in the plane}
generally exhibit a much more interesting behavior than models {\it
on the line}; this is conf\/irmed by the results reported below, as
well as by those displayed in previous papers [1]--[4].

\setcounter{equation}{0}
\renewcommand{\theequation}{\arabic{section}.\arabic{equation}}

\section{The solution}

In this Section we indicated how to solve the system (1.1).

Let the $n$ complex quantities $f_j(t)$ evolve in time according to
the {\it linear firstorder} evolution equations
\be
\dot{f}_j=\sum_{k=1}^n \alpha_{jk} f_k,
\ee
whose general solution reads
\be
f_j(t)=\sum_{k}^n \varphi^{(k)}_j \exp(a_k t),
\ee
where the $n$ quantities $a_k$ are the $n$ eigenvalues of the
$(n\times n)$-matrix $A$ with elements $\alpha_{jk}$, and the $n$
$n$-vectors $\Phi^{(k)}$, with components $\varphi^{(k)}$, are the
corresponding eigenvectors:
\be
\sum_{k=1}^n\alpha_{jk} \varphi^{(m)}_k =a_m \varphi^{(m)}_j.
\ee
Here we are implicitly assuming, for simplicity, that the matrix $A$
is diagonalizable, and that all its eigenvalues $a_k$ are distinct;
the diligent reader will have no dif\/f\/iculty in detailing the
modif\/ications in the treatment which are required if these
assumptions do not hold.

Introduce now the $n$ quantities $z_j(t)$ via the position
\be
f_j=\dot z_j/z_j,
\ee
which clearly entails
\be
\dot f_j=\ddot z_j/z_j -\dot z_j^2/z_j^2,
\ee
hence, via (2.1) and (2.4),
\be
\ddot z_j=\dot z_j^2/z_j +\sum_{k=1}^n \alpha_{jk} z_j \dot z_k/z_k.
\ee

On the other hand, from (2.4) and (2.2), one gets
\be
z_j(t) =z_j(0)\exp\left[ \sum_{k=1}^n \varphi^{(k)}_j \left[ \exp(a_k
t)-1\right] /a_k\right],
\ee
while the normalization of the $n$ $n$-vectors $\Phi^{(k)}$, with
components $\varphi^{(k)}_j$, is f\/ixed by the conditions
\be
\sum_{k=1}^n \varphi^{(k)}_j =\dot z_j (0)/z_j(0),
\ee
also implied by (2.2) and (2.4).

Hence (2.6) is explicitly solvable, for any choice of the $n^2$
quantities $\alpha_{jk}$, and $2n$ initial data $z_j(0)$ and $\dot
z_j(0)$. And this is true as well if all these quantities are {\it
complex}, so that we can put
\be
z_j=x_j+i y_j,
\ee
\be
\alpha_{jk}=\beta_{jk}+i \gamma_{jk},
\ee
of course now with $x_j$, $y_j$, $\beta_{jk}$ and $\gamma_{jk}$ all
{\it real}.

But it is now a trivial exercise to verify that, via these positions,
the {\it complex} evolution equations (2.6) coincide with the {\it
$2$-dimensional rotation-invariant real} equations of motion~(1.1),
whose solvability is thereby demonstrated.

\setcounter{equation}{0}

\section{Discussion  of the motions, and other models}

In this Section we brief\/ly discuss the motions yielded by the
many-body problem (1.1), using the f\/indings of the preceding
Section(see in particular (2.7)); and we then indicate how to obtain
other solvable models {\it in the plane}, including the many-body
problem (1.3) and {\it translation-invariant} generalizations of the
many-body problems mentioned above.

Clearly the most important element that characterizes the behavior of
the system (1.1) are the $n$ eigenvalues of the matrix $A$. If {\it
all} these eigenvalues have negative real parts, from {\it any}
initial conditions the system will tend to a standstill
conf\/iguration; {\it all} particle velocities vanish exponentially as
$t\to \infty$. Such an outcome may also occur, but only for {\it
special} initial conditions, if only {\it some} of these eigenvalues
have negative real parts. A necessary and suf\/f\/icient condition for the
system to possess periodic trajectories is that {\it at least one} of
these $n$ eigenvalues be imaginary. If {\it all} these eigenvalues
are imaginary, the motion remains conf\/ined, and is in fact multiply
periodic, for {\it any} initial conditions. If moreover {\it all}
these eigenvalues are rational multiples of the {\it same} imaginary
quantity, then for {\it any} initial conditions the system behaves
completely periodically. If one of these eigenvalues vanishes, the
system (2.6) possesses the {\it similarity solutions}
\setcounter{equation}{0}
\renewcommand{\theequation}{\arabic{section}.\arabic{equation}{\rm a}}
\be
z_j(t)=z_j(0)\exp(\eta t),
\ee
with $\eta =\lambda +i\omega$ an arbitrary (complex) constant. The
corresponding {\it similarity solution} of~(1.1) reads
\setcounter{equation}{0}
\renewcommand{\theequation}{\arabic{section}.\arabic{equation}{\rm b}}
\be
\vec r_j(t) =\exp(\lambda t) \left[ \cos(\omega t) +\sin(\omega
t)\hat z \wedge\right] \vec r_j(0).
\ee
\setcounter{equation}{1}
\renewcommand{\theequation}{\arabic{section}.\arabic{equation}}
If some of the $n$ eigenvalues $a_k$ have positive real parts, the
generic solution may feature {\it some} (from 0 to $n$) particles
which escape, doubly exponentially fast, to inf\/inity, while the
others converge, doubly exponentially fast, to the origin.

Let us now derive the system (1.3a) with (1.3b). To this end let us
start from the system~(2.6), but with the quantities $z_j(t)$
formally replaced by, say, $\zeta_j(\tau)$ (and accordingly with dots
replaced by, say, primes, the latter indicating derivatives with
respect to $\tau$). We then introduce new quantities $z_j(t)$ via the
position $z_j(t)=\zeta_j(\tau)$, so that (2.6) gets replaced by
\be
\ddot z_j=\dot z_j(\ddot \tau/ \dot \tau)+\dot z_j^2/z_j+\dot \tau
\sum_{k=1}^n \alpha_{jk} z_j \dot z_k/z_k.
\ee
Now we make again the transition from the {\it complex} dependent
variables $z_j(t)$ to the {\it real $2$-vectors} $\vec r_j(t)$ (as
above, see (2.9) and (1.2)). It is then easily seen that the new
particle coordinates $\vec r_j(t)$ satisfy precisely (1.3a) with
(1.3b), provided
\be
\tau=\left\{\exp[(\lambda +i\omega)t]-1\right\}/(\lambda+i\omega).
\ee
The statement made after (1.3b) is thereby proven.

Finally, let us manufacture a generalized version of (1.1), which is
{\it translation-invariant}. To this end we introduce a model with $2n$
particles of two types, whose coordinates we denote respectively by
$\vec r^{\;(+)}_j(t)$ and $\vec r^{\;(-)}_j(t)$, and we set
\be
\vec r_j(t)=\vec r^{\;(+)}_j(t)-\vec r^{\;(-)}_j(t), \qquad
\vec R_j(t)=\vec r^{\;(+)}_j(t)+\vec r^{\;(-)}_j(t),
\ee
with $\vec r_j(t)$ satisfying the (solvable) equations of motion
(1.1), and $\vec R_j(t)$ satisfying the (clearly, also solvable; see
below) equations of motion
\be
\ddot{\vec R_j}=(\Lambda_j+\Omega_j\hat z\wedge)\dot{\vec R_j}.
\ee
Here $\lambda_j$ and $\Omega_j$ are $2n$ arbitrary (of course {\it
real}) coupling constants.

We thereby see that the new ``particle coordinates'' $\vec
r^{(\pm)}_j(t)$ evolve according to the {\it translation-invariant}
equations of motion
\be
\ba{l}
\ds
{\ddot{\vec r}}_j^{\;(\pm)}=\frac 12 \Bigl\{(\Lambda_j+\Omega_j\hat
z\wedge) \dot{\vec R_j}\pm
\frac{ 2\dot{\vec r_j}(\dot{\vec r_j}\cdot {\vec
r_j}) -{\vec r_j}(\dot{\vec r_j}\cdot \dot {\vec r_j})}{r_j^2}\\[5mm]
\ds \qquad \qquad \qquad \pm \sum_{k=1}^n (\beta_{jk}+\gamma_{jk}
\hat{z} \wedge) \frac{\dot{\vec r_k}({\vec r_k}\cdot {\vec r_j}) +
{\vec r_j}(\dot{\vec r_k}\cdot {\vec r_k})-
{\vec r_k}(\dot{\vec r_k}\cdot {\vec r_j})}{r_k^2}\Bigr\},
\ea
\ee
of course  with (3.4).

As for (3.5), a convenient way to exhibit its solution is via the
formula
\be
Z_j(t)=Z_j(0)+\dot
Z_j(0)\left\{\exp\left[(\Lambda_j+i\Omega_j)t\right]\right\} /
(\Lambda_j +i \Omega_j),
\ee
which clearly provides the general solution of (3.5) via the
identif\/ication of the {\it complex} dependent variables $Z_j\equiv
X_j+i Y_j$, with the {\it real} 2-vectors $\vec R_j\equiv(X_j,\, Y_j)$,
since this identif\/ication entails that, to (3.5), there corresponds
\be
\ddot Z_j=(\Lambda_j+i\Omega_j)\dot Z_j.
\ee

Clearly the same trick may be directly applied to (1.3a) with (1.3b),
yielding a {\it tran\-sla\-tion-in\-va\-riant} generalization of these
equations of motion. An alternative possibility, generally yielding a
dif\/ferent model, is to apply the change of independent variable (3.3)
to the complex scalar version of (3.6), and {\it subsequently} to
change the dependent variables (from scalar complex numbers to real
2-vectors). We do not exhibit the corresponding results, which the
interested reader may immediately write down. Nor do we report an
analysis of the actual behavior of the solutions of (3.6) with (3.4),
or of the new models just mentioned, since the above results are so
transparent not to require elaborations. But we end by reemphasizing
the reachness of the motions entailed by these formulas, which is now
enhanced by the additional freedom connected with the choice of the
$2n$ constants $\Lambda_j$ and $\Omega_j$.
\label{calogero-lp}

\end{document}